\documentclass[twocolumn,showpacs,preprintnumbers,amsmath,amssymb]{revtex4}

\usepackage{graphicx}
\usepackage{dcolumn}
\usepackage{bm}
\usepackage{graphics}
\usepackage{amsmath}
\usepackage{amssymb}
\usepackage{amscd}
\usepackage{afterpage}
\usepackage{float,times}
\usepackage{subfigure}
\usepackage{rotating}
\usepackage{multirow}
\usepackage{fancyheadings}
\usepackage{epsfig}
\usepackage{theorem}
\usepackage{moreverb}
\usepackage{euscript}
\usepackage{psfrag}

\begin{document}

\title{Quark-Lepton Symmetry In Five Dimensions}
\author{A. Coulthurst}\email{a.coulthurst@physics.unimelb.edu.au}
\author{K. L. McDonald}\email{k.mcdonald@physics.unimelb.edu.au}

\author{B. H. J. McKellar}
 \email{b.mckellar@physics.unimelb.edu.au}
\affiliation{
School of Physics, Research Centre for High Energy Physics, The
University of Melbourne, Victoria, 3010, Australia\\
}


\begin{abstract}
We construct a complete five dimensional Quark-Lepton symmetric
model, with all fields propagating in the bulk. The extra dimension
forms an $S^1/Z_2\times Z_2'$ orbifold with the zero mode fermions
corresponding to standard model quarks localised at one fixed
point. Zero modes corresponding to left(right)-chiral leptons are
localised at (near) the other fixed point. This localisation pattern
is motivated by the symmetries of the model. Shifting the right-handed
neutrinos and charged leptons slightly from the fixed point provides a new
mechanism for understanding the absence of
relations of the type $m_e=m_u$ or $m_e=m_d$ in Quark-Lepton
symmetric models. Flavour changing neutral currents resulting from Kaluza Klein
gluon exchange, which typically arise in the quark sector of split fermion
models, are suppressed due to the localisation of quarks at one
point. The separation of quarks and
leptons in the compact extra
dimension also acts to suppress the proton decay rate. This permits
the extra dimension to be much larger than that obtained in a previous
construct, with the bound $1/R\gtrsim30$~TeV obtained.
\end{abstract}
\maketitle
\section{Introduction\label{sec:bulk_5d_ql_intro}}
It is known that the evident differences between quarks and leptons
may be no more than the low energy manifestation of a more symmetric
underlying theory. In particular it has been shown that one may
construct four dimensional models which are invariant under a discrete
quark-lepton (QL) symmetry, whereby one interchanges all quarks and
leptons in the
Lagrangian~\cite{Foot:1990dw,Foot:1990um,Foot:1990un,Foot:1991fk,Levin:1993sq,Shaw:1994zs,Foot:1995xx}.
This idea requires the introduction of leptonic colour, an $SU(3)$
gauge group acting on a set of generalised leptons, and thus predicts
new gauge bosons and fermions.

Recently the notion of a QL symmetry has been investigated in five
dimensions~\cite{McDonald:2006dy}. In that work the scalar and gauge
fields of the QL
symmetric model were assumed to propagate in the bulk, whilst all quark
and lepton fields were confined to a brane. The extra dimension was
assumed to form an $S^1/Z_2\times Z_2'$ orbifold, the construction of
which reduced the five dimensional QL symmetric gauge group to one of
its subgroups. The model required an order $10^{11}$~GeV cut-off to
suppress the proton decay rate and the extra dimension was taken to be
of order $1/R\sim 10^{9}$~GeV.

In the present work we construct a complete five
dimensional QL
symmetric model, with all fields assumed to propagate in a five
dimensional spacetime. Placing fermions in the bulk
provides new ways of addressing the problems encountered in the
previous construct. Provided that the Standard Model (SM) quarks and
leptons are localised at (or near) different ends of the compact extra
dimension it may be as large as $1/R\sim 30$~TeV,
permitting the observation of higher dimensional physics at future
colliders. 

A common problem encountered in QL symmetric models is the presence of
tree level mass
relations of the type $m_e=m_u$ or $m_e=m_d$. The higher dimensional
framework provides a new way to understand the absence of such mass
relations. The mass relations may be removed from the effective four
dimensional theory if zero mode fermions corresponding to SM quarks and
leptons have different profiles along
the extra dimension. In our model this occurs as follows.
All zero mode fermions corresponding to SM quarks are localised at one
orbifold fixed point, with
the varying widths of the fifth dimensional quark wavefunctions
determining the degree of wavefunction overlap between left- and
right-chiral SM
quarks. This in turn determines the size of the effective four dimensional quark Yukawa
couplings with the SM Higgs scalar. Leptons are
localised at the opposite end of the extra dimension, with $SU_L(2)$ doublet
leptons localised at the fixed point and right-chiral
neutrinos and charged leptons shifted slightly into the bulk. Thus
the fifth dimensional wavefunction overlaps in the lepton sector tend
to be reduced relative to that of the quark sector, motivating the
flavour differences observed between quarks and leptons. By shifting
the right-chiral neutrinos further into the bulk than the right-chiral
charged leptons, neutrino Dirac masses may be suppressed below the
electroweak scale.

The symmetries of the model motivate the localisation pattern of SM
fermions outlined above. The localisation of quarks at
one point in the extra dimension allows one to evade the flavour changing neutral current
bounds on the size of the extra dimension which arise generically in the quark sector of split fermion
models~\cite{Lillie:2003sq}. These bounds may be quite severe, requiring $1/R$ to be as
large as $5000~\mathrm{TeV}^{-1}$~\cite{Delgado:1999sv}. The flavour
changing neutral
current bounds which arise in the lepton sector are much weaker and
permit the extra dimension to be relatively large.

We note that the concept of leptonic colour has
recently been generalised in~\cite{Foot:2006ie} and studied within the
context of unified theories
in~\cite{Joshi:1991yn,Babu:2003nw,Chen:2004jz,Demaria:2005gk,Demaria:2006uu,Demaria:2006bd}.
Split fermions have also been employed to
remove quark-lepton mass relations in a different context
in~\cite{Hung:2004ac}.

The layout of this paper is as follows. In
Section~\ref{sec:bulk_5d_ql_gauge_sector} the gauge and scalar content
of the model are
discussed. Section~\ref{sec:bulk_5d_ql_fermion_parities} introduces
fermions and we determine the orbifold parity assignments necessary
to obtain a realistic zero mode fermion spectrum. The Yukawa sector of
the model is introduced in Section~\ref{sec:bulk_5d_ql_yukawas} whilst
the issue of proton decay is addressed in
Section~\ref{sec:bulk_5d_ql_p_decay}. Fermion masses are considered in
Section~\ref{sec:bulk_5d_ql_fermion_mass}, with the neutrino
sector considered separately in Section~\ref{bulk_ql_nu_mass}. The
size of the extra dimension is discussed in
Section~\ref{sec:rev_bulk_ql_bounding_r} whilst
Section~\ref{sec:bulk_5d_ql_sterile_cosmology} addresses the lifetime
and cosmological bounds on
the mass of the lightest SM singlet neutrino. A discussion of the
expected scale for new physics and some experimental signatures is provided in
Section~\ref{sec:bulk_ql_bounding_scales} and the paper concludes in
Section~\ref{sec:bulk_5d_ql_conc}.
\section{The Gauge and Scalar Sectors\label{sec:bulk_5d_ql_gauge_sector}}
The QL symmetric gauge group is
\begin{eqnarray}
\mathcal{G}_{ql}=SU_l(3)\otimes SU_c(3)\otimes SU_L(2)\otimes U_X(1),
\end{eqnarray}
where $SU_l(3)$ is the lepton colour group and $SU_c(3)$ is the usual colour
group for quarks. The action of the group $SU_L(2)$ on the zero mode
fermion spectrum will be identified
with the usual weak group of the
SM and $X\ne Y$, where $Y$ is the SM hypercharge.
Under the discrete QL symmetry the gauge fields transform as:
\begin{eqnarray}
G_c^M\leftrightarrow G_l^M,\mkern20mu W^M\leftrightarrow W^M,\mkern20mu C^M\leftrightarrow -C^M,
\end{eqnarray}
where $G_{c,l}^M$ are the $SU_{c,l}(3)$ gauge bosons, $W^M$ are
the weak bosons and $C^M$ is the $U_X(1)$ gauge boson. The five
dimensional Lorentz
index takes the values $M=\mu,5$, where $\mu$ is the $3+1$ dimensional
Lorentz index.  The additional spatial dimension is taken as the
orbifold $S^1/Z_2\times Z_2'$, whose coordinate is labelled as
$y$. The construction of the orbifold proceeds via the identification
$y\rightarrow -y$ under $Z_2$ and $y'\rightarrow-y'$
under the $Z_2'$ symmetry, where $y'=y+\pi R/2$. The physical region
in $y$ is given by the interval $[0,\pi R/2]$. The Lagrangian is
required to be invariant under the discrete $Z_2\times Z_2'$ symmetry,
whose action in the space gauge fields is defined as follows:
\begin{eqnarray}
W_\mu(x^\mu,y)&\rightarrow& W_\mu(x^\mu,-y)=PW_\mu(x^\mu,y)P^{-1},\nonumber\\
W_5(x^\mu,y)&\rightarrow& W_5(x^\mu,-y)=-PW_5(x^\mu,y)P^{-1},\nonumber\\
W_\mu(x^\mu,y')&\rightarrow& W_\mu(x^\mu,-y')=P'W_\mu(x^\mu,y')P'^{-1},\nonumber\\
W_5(x^\mu,y')&\rightarrow& W_5(x^\mu,-y')=-P'W_5(x^\mu,y')P'^{-1},\nonumber
\end{eqnarray}
where $W$ denotes a generic gauge field.
We take $P$ and $P'$ to be trivial for the $SU_c(3)$,
$SU_L(2)$ and $U_X(1)$ gauge bosons. For the $SU_l(3)$ gauge bosons we
choose $P=\mathrm{diag}(1,1,1)$ and $P'=\mathrm{diag}(-1,1,1)$, which
will reduce the leptonic colour symmetry to $SU_l(2)\otimes U_{X'}(1)$
in the four dimensional effective theory. We write the five dimensional $SU_l(3)$ gauge bosons as
\begin{eqnarray}
G_l&=&T_a G_l^a \nonumber\\
&=&\left( \begin{array}{ccc}
           -\frac{2}{\sqrt{3}}G^8&\sqrt{2}Y^1 &\sqrt{2}Y^2    \\
           \sqrt{2}Y^{1\dagger}&G^3+\frac{1}{\sqrt{3}}G^8& \sqrt{2}\tilde{G}\\
           \sqrt{2}Y^{2\dagger}&\sqrt{2}\tilde{G}^\dagger&-G^3+\frac{1}{\sqrt{3}}G^8  
\end{array} \right),
\end{eqnarray}
and find their $Z_2\times Z_2'$ parities to be
\begin{eqnarray}
Y_{\mu}^1,Y_{\mu}^2,Y_{\mu}^{1\dagger},Y_{\mu}^{2\dagger}&\rightarrow& (+,-),\nonumber\\
Y_{5}^1,Y_{5}^2,Y_{5}^{1\dagger},Y_{5}^{2\dagger}&\rightarrow& (-,+),\nonumber\\
G^8_\mu,G^3_\mu,\tilde{G}_\mu,\tilde{G}_\mu^\dagger&\rightarrow& (+,+),\nonumber\\
G^8_5,G^3_5,\tilde{G}_5,\tilde{G}_5^\dagger&\rightarrow& (-,-).
\end{eqnarray}
One may expand the gauge bosons as a Fourier series in the compact
extra dimension, with the $Z_2\times Z_2'$ parities constraining the
series as usual. One has
\begin{widetext}
\begin{eqnarray}
\psi_{(+,+)}(x^\mu,y)&=&\frac{2}{\sqrt{\pi R}}\left(\frac{1}{\sqrt{2}}\psi^{(0)}_{(+,+)}(x^\mu)+\sum_{n=1}^\infty\psi^{(n)}_{(+,+)}(x^\mu)\cos\frac{2ny}{R}\right),\nonumber\\
\psi_{(+,-)}(x^\mu,y)&=&\frac{2}{\sqrt{\pi R}}\sum_{n=1}^\infty\psi^{(n)}_{(+,-)}(x^\mu)\cos\frac{(2n-1)y}{R},\nonumber\\
\psi_{(-,+)}(x^\mu,y)&=&\frac{2}{\sqrt{\pi R}}\sum_{n=1}^\infty\psi^{(n)}_{(-,+)}(x^\mu)\sin\frac{(2n-1)y}{R},\nonumber\\
\psi_{(-,-)}(x^\mu,y)&=&\frac{2}{\sqrt{\pi R}}\sum_{n=1}^\infty\psi^{(n)}_{(-,-)}(x^\mu)\sin\frac{2ny}{R},\label{bulk_5d_ql_fourier_expansion}
\end{eqnarray}
\end{widetext}
where $\psi$ represents a generic field. Thus the four dimensional
charge 1/2 bosons $Y_{\mu}^1$, $Y_{\mu}^2$, $Y_{\mu}^{1\dagger}$ and
$Y_{\mu}^{2\dagger}$ do not possess zero modes, with the $n$th mode
possessing a mass of $(2n-1)/R$. The fields $G^8_\mu$, $G^3_\mu$,
$\tilde{G}_\mu$ and $\tilde{G}_\mu^\dagger$ all have zero modes, with
the higher modes possessing a mass of $2n/R$. After compactification
the zero mode gauge group is
\begin{eqnarray}
\mathcal{G}_{ql}\rightarrow SU_l(2)\otimes SU_c(3)\otimes SU_L(2)\otimes U_{X'}(1)\otimes U_X(1),\label{bulk_5d_ql_compact_gauge_group}
\end{eqnarray}
and the next stage of symmetry breaking requires
\begin{eqnarray}
U_{X'}(1)\otimes U_X(1)\rightarrow U_Y(1),
\end{eqnarray}
which shall be achieved by the usual Higgs mechanism. The scalar
content necessary to break $\mathcal{G}_{ql}$ is
\begin{eqnarray}
& &\chi\sim(3,1,1,2/3),\mkern15mu
\chi'\sim(1,3,1,2/3),\nonumber\\
& &\mkern20mu\mathrm{and}\mkern20mu \phi\sim(1,1,2,1),
\end{eqnarray}
with the action of the discrete QL symmetry given by,
\begin{eqnarray}
\phi\leftrightarrow \tilde{\phi},\mkern20mu \chi\leftrightarrow \chi',
\end{eqnarray}
where $\tilde{\phi}=\epsilon \phi^*$ and $\epsilon$ is the two
dimensional anti-symmetric tensor. The $Z_2\times
Z_2'$ parities of $\phi$ are trivial whilst
for $\chi$ we have
\begin{eqnarray}
\chi(x^\mu,y)&\rightarrow& \chi(x^\mu,-y)=P\chi(x^\mu,y),\nonumber\\
\chi(x^\mu,y')&\rightarrow&
\chi(x^\mu,-y')=-P'\chi(x^\mu,y'),\label{bulk_5d_ql_intro_chi_parities}
\end{eqnarray}
with $P=\mathrm{diag}(1,1,1)$ and $P'=\mathrm{diag}(-1,1,1)$. For $\chi'$ we take
\begin{eqnarray}
\chi'(x^\mu,y)&\rightarrow& \chi'(x^\mu,-y)=P\chi'(x^\mu,y),\nonumber\\
\chi'(x^\mu,y')&\rightarrow& \chi'(x^\mu,-y')=P'\chi'(x^\mu,y'),\label{bulk_5d_ql_intro_chi'_parities}
\end{eqnarray}
with $P$ and $P'$ trivial. Under the symmetry reduction
\begin{eqnarray}
SU_l(3)\rightarrow SU_l(2)\otimes U_{X'}(1),
\end{eqnarray} 
one has
\begin{eqnarray}
\chi\rightarrow \chi_2 \oplus \chi_1,
\end{eqnarray}
where $\chi_2\sim(2,1)$ and $\chi_1\sim(1,-2)$ have the $Z_2\times Z_2'$ parities
\begin{eqnarray}
\chi_1\rightarrow (+,+)\mkern15mu,\mkern15mu\chi_2\rightarrow(+,-).
\end{eqnarray}
The zero mode for $\chi_1$ may develop a VEV to break the gauge
symmetry as follows:
\begin{eqnarray}
& &SU_l(2)\otimes SU_c(3)\otimes SU_L(2)\otimes U_{X'}(1)\otimes
U_X(1)\nonumber\\
& &\mkern80mu\downarrow\nonumber\\
& &SU_l(2)\otimes SU_c(3)\otimes SU_L(2)\otimes U_Y(1).
\end{eqnarray}
At this stage the hypercharge generator may be identified as
\begin{eqnarray}
Y=X+\frac{1}{\sqrt{3}}T_8,
\end{eqnarray}
where $T_8=(1/\sqrt{3})\times\mathrm{diag}(-2,1,1)$ is a diagonal
generator of $SU_l(3)$. The final stage of symmetry breaking occurs
when the neutral
component of $\phi$ develops the VEV $u$ to give
\begin{eqnarray}
& &SU_l(2)\otimes SU_c(3)\otimes SU_L(2)\otimes U_Y(1)\nonumber\\
& &\mkern80mu\downarrow\nonumber\\
& &SU_l(2)\otimes SU_c(3)\otimes U_Q(1).
\end{eqnarray}
Assuming $g_S^2\gg g_X^2,g_L^2$, where $g_L$ [$g_X$] is the $SU_L(2)$ [$U_X(1)$] coupling constant and
$g_S$ denotes the common $SU_l(3)$ and $SU_c(3)$ coupling constant,
one may write the neutral gauge boson mass eigenvalues as~\cite{McDonald:2006dy}
\begin{eqnarray}
M^{2(n)}_{\gamma}&=&\left(\frac{2n}{R}\right)^2,\nonumber\\
M_{Z}^{2(n)}&\simeq&\frac{1}{2}(g_X^2+g_L^2)u^2-\frac{g_X^2}{6g_S^2}u^2+\left(\frac{2n}{R}\right)^2,\nonumber\\
M_{Z'}^{2(n)}&\simeq&\frac{2}{3}g_S^2w^2\left\{1+\frac{g_X^2}{3g_S^2}\right\}+\left(\frac{2n}{R}\right)^2.
\end{eqnarray}
Note that the zero modes possess the same masses as the neutral
gauge bosons in the minimal four dimensional QL symmetric
model~\cite{Foot:1990dw,Foot:1991fk}. These zero modes couple
to fermions in exactly the same way as the neutral gauge bosons in the
minimal QL symmetric model, making the phenomenology of these sates
identical to that of the neutral gauge bosons studied
in~\cite{Foot:1991fk}.

The zero modes consist of the massless photon, the $Z$ boson with mass
of order $u$, the electroweak scale, and an additional neutral boson
$Z'$ with mass of order $w$, the $U_{X'}(1)\otimes U_X(1)$ symmetry
breaking scale. The phenomenological bound of $M_{Z'}>720$~GeV
obtained in~\cite{Foot:1991fk} also applies to the zero mode $Z'$
boson in the present model. Thus we obtain a lower bound on the
$U_{X'}(1)\otimes U_X(1)$ symmetry breaking scale of $w\gtrsim1$~TeV,
which is low enough to permit observation of the $Z'$ boson at the
LHC.

The
Kaluza-Klein (KK) tower
of charged 1/2 bosons possess the mass
\begin{eqnarray}
M_{Y^1}^{2(n)}=M_{Y^2}^{2(n)}=\frac{1}{2}g_S^2w^2+\left(\frac{2n-1}{R}\right)^2,\end{eqnarray}
where the zero mode is absent. We shall discuss the bounds on $R$ and
the $Y$ boson mass scales in Section~\ref{sec:rev_bulk_ql_bounding_r}. 

The mass of the $W$ bosons are given by
\begin{eqnarray}
M^{2(n)}_W=\frac{1}{2}g_L^2u^2 +\left(\frac{2n}{R}\right)^2,
\end{eqnarray}
with the zero mode corresponding to the usual $W$ boson.
\section{Fermion Parities\label{sec:bulk_5d_ql_fermion_parities}}
 Whilst the five
dimensional theory is vector-like and necessarily free from
anomalies, we shall be introducing four dimensional chirality via the
orbifold boundary conditions~\cite{Georgi:2000wb}. The effective four
dimensional gauge theory is free from anomalies if all anomalies
cancel amongst the zero mode fermions~\cite{Arkani-Hamed:2001is}. After compactification the
gauge group is given in equation (\ref{bulk_5d_ql_compact_gauge_group}).
To cancel anomalies involving $U_{X'}(1)$ factors one requires both the SM
leptons and
extra coloured leptons to appear at the zero mode level. This may be
achieved by doubling the fermion content of the QL symmetric model, in
analogy with the doubling required in five dimensional left-right symmetric
models~\cite{Mohapatra:2002rn,Perez:2002wb}. We denote the fermions by
\begin{eqnarray}
Q_i&\sim& (1,3,2,1/3),\mkern20mu L_i\sim(3,1,2,-1/3),\nonumber\\
U^c_i&\sim&(1,\bar{3},1,-4/3),\mkern20mu E^c_i\sim(\bar{3},1,1,4/3),\nonumber\\
D^c_i&\sim&(1,\bar{3},1,2/3),\mkern20mu N^c_i\sim(\bar{3},1,1,-2/3),
\end{eqnarray}
where the quantum numbers under $\mathcal{G}_{ql}$ are shown and
$i=1,2$ labels distinct fermion multiplets. The action of the QL
symmetry on the fermions is
\begin{eqnarray}
Q_i\leftrightarrow L_i,\mkern20mu U^c_i\leftrightarrow
E^c_i,\mkern20mu D^c_i\leftrightarrow N^c_i.
\end{eqnarray}
The transformation rules for a fermion field under the $Z_2\otimes
Z_2'$ discrete symmetries are
\begin{eqnarray}
F(x^\mu,y)&\rightarrow&F(x^\mu,-y)=\pm\gamma_5PF(x^\mu,y),\nonumber\\
F(x^\mu,y')&\rightarrow&F(x^\mu,-y')=\pm\gamma_5P'F(x^\mu,y'),
\end{eqnarray}
where $F$ denotes a generic fermion field and the signs may be chosen
independently for the distinct discrete symmetries. The matrices
$P$ and $P'$ are identical to the ones used for the gauge sector in
the last section, with the only non-trivial matrix being $P'$ acting in
leptonic colour space. We demand that the lepton fields acquire the $Z_2\otimes
Z_2'$ quantum numbers:
\begin{eqnarray*}
& & L_{1}=\begin{pmatrix}L^1_{1L}(+,+)\cr
              L^2_{1L}(+,-)\cr L^3_{1L}(+,-)\cr L^1_{1R}(-,-)\cr
              L^2_{1R}(-,+)\cr L^3_{1R}(-,+) \end{pmatrix}, \mkern25mu L_2=\begin{pmatrix}L^1_{2L}(+,-)\cr
              L^2_{2L}(+,+)\cr L^3_{2L}(+,+)\cr L^1_{2R}(-,+)\cr
              L^2_{2R}(-,-)\cr L^3_{2R}(-,-) \end{pmatrix}, \\
& & N^c_{1}=\begin{pmatrix}n^{1c}_{1L}(+,+)\cr
              n^{2c}_{1L}(+,-)\cr n^{3c}_{1L}(+,-)\cr n^{1c}_{1R}(-,-)\cr
              n^{2c}_{1R}(-,+)\cr n^{3c}_{1R}(-,+) \end{pmatrix},
            \mkern25mu  N^c_{2}=\begin{pmatrix}n^{1c}_{2L}(+,-)\cr
              n^{2c}_{2L}(+,+)\cr n^{3c}_{2L}(+,+)\cr n^{1c}_{2R}(-,+)\cr
              n^{2c}_{2R}(-,-)\cr n^{3c}_{1R}(-,-) \end{pmatrix},\\
& &E^c_{1}=\begin{pmatrix}e^{1c}_{1L}(+,+)\cr
              e^{2c}_{1L}(+,-)\cr e^{3c}_{1L}(+,-)\cr e^{1c}_{1R}(-,-)\cr
              e^{2c}_{1R}(-,+)\cr e^{3c}_{1R}(-,+)
            \end{pmatrix},\mkern25mu E^c_{2}=\begin{pmatrix}e^{1c}_{2L}(+,-)\cr
              e^{2c}_{2L}(+,+)\cr e^{3c}_{2L}(+,+)\cr e^{1c}_{2R}(-,+)\cr
              e^{2c}_{2R}(-,-)\cr e^{3c}_{2R}(-,-)
            \end{pmatrix}.  
\end{eqnarray*}
Here the numerical superscripts $1,2,3$ label the different lepton
colours and the numerical subscripts $1,2$ label the different five
dimensional fields. As $SU_c(3)$ is not broken by the orbifold
compactification the three quark colours all possess the same
$Z_2\otimes Z_2'$ parities. Suppressing the quark colour index,
we enforce the following orbifold parities for quarks:

\begin{eqnarray*}
&
&Q_1=\begin{pmatrix}Q_{1L}(+,+)\cr Q_{1R}(-,-)\end{pmatrix},\mkern25mu
Q_2=\begin{pmatrix}Q_{2L}(+,-)\cr Q_{2R}(-,+)\end{pmatrix},
\\
&
&U^c_1=\begin{pmatrix}u^c_{1L}(+,+)\cr
  u^c_{1R}(-,-)\end{pmatrix},\mkern25mu
U^c_2=\begin{pmatrix}u^c_{2L}(+,-)\cr u^c_{2R}(-,+)\end{pmatrix},
\\
&
&D^c_1=\begin{pmatrix}d^c_{1L}(+,+)\cr
  d^c_{1R}(-,-)\end{pmatrix},\mkern25mu D^c_2=\begin{pmatrix}d^c_{2L}(+,-)\cr d^c_{2R}(-,+)\end{pmatrix}.
\end{eqnarray*}

Only fermion fields with the orbifold parities $(+,+)$ appear at
the
zero mode level. We identify the zero modes of the fields $L^1_{1L}$,
$e^{1c}_{1L}$, $Q_{1L}$, $u_{1L}^c$ and $d_{1L}^c$ with the SM fields
$L$, $e_R^c$, $Q_L$, $u_R^c$ and $d_R^c$ respectively. The zero modes
of the fields
$L^{2,3}_{2L}$, $e^{2,3c}_{2L}$ and $n^{2,3c}_{2L}$ are the usual
exotic leptons found in QL symmetric models (known as
liptons in the literature~\cite{Foot:1991fk}). The liptons form
doublets under the remnant leptonic
colour symmetry $SU_l(2)\subset SU_l(3)$ and are confined into two
particle bound states. These states form one of the key signatures of
QL symmetric models, leading to an exotic spectrum of particles which
may decay into SM fields by creation of electroweak gauge
bosons~\cite{Foot:2006ie,McDonald:2006dy}. We
shall see below that the
zero mode liptons acquire mass at the symmetry breaking scale $w$ and
may be observed at the LHC. Note that the zero mode liptons appear in
different multiplets to the SM leptons. Consequently the lightest liptons will
not couple directly to SM leptons via $Y$-boson exchange. This is one
of the major
phenomenological differences between the current model and that
constructed in~\cite{McDonald:2006dy}. We explore the implications of
this difference below.

The zero mode of $n^{1c}_{1L}$
is an additional neutrino which is sterile under the electroweak gauge
group. This state will play the role of the usual SM singlet neutrino
and we shall label it as $\nu_R^c$. We shall discuss the mass of
this state in
Section~\ref{bulk_ql_nu_mass} and issues relating to its cosmological
evolution in Section~\ref{sec:bulk_5d_ql_sterile_cosmology}.
\section{Yukawa Couplings\label{sec:bulk_5d_ql_yukawas}}
The Yukawa Lagrangian must remain invariant under both gauge and orbifold
parity transformations. It is convenient to split the five dimensional
Yukawa Lagrangian into two portions, with the non-electroweak portion
given by
\begin{widetext}
\begin{eqnarray}
\mathcal{L}_{non-ew}&=&\frac{1}{\sqrt{\Lambda}}\left\{h_1[L_1^TC_5^{-1}L_1\chi +Q_1^TC_5^{-1}Q_1\chi']+h_2[L_2^TC_5^{-1}L_2\chi +Q_2^TC_5^{-1}Q_2\chi']+\right.\nonumber\\
& &\left.h_1'[N_1^{cT}C_5^{-1}E_1^c\chi^\dagger
+D_1^{cT}C_5^{-1}U_1^c\chi'^\dagger]+h_2'[N_2^{cT}C_5^{-1}E_2^c\chi^\dagger
+D_2^{cT}C_5^{-1}U_2^c\chi'^\dagger]+\mathrm{H.c.}\right\}.\label{bulk_5d_ql_non-ew_lagrangian}
\end{eqnarray}
\end{widetext}
Here the $h$'s are Yukawa coupling
matrices in flavour space, $C_5=\gamma^0\gamma^2\gamma^5$ is the
five dimensional charge conjugation matrix and $\Lambda$ is the cut-off. Contraction over
$SU_{c,l}(3)$ colour
indices with the
three dimensional anti-symmetric tensor
$\epsilon^{\alpha\beta\gamma}$ is implied in
(\ref{bulk_5d_ql_non-ew_lagrangian}). When the zero mode of $\chi$
develops a VEV \mbox{$\langle \chi^{(0)}\rangle=w$} the $h_2$ and $h_2'$ terms in
(\ref{bulk_5d_ql_non-ew_lagrangian}) generate an order $w$ mass for
the liptons. All SM fields remain massless at this stage of symmetry breaking.

The electroweak portion of the Yukawa Lagrangian is
\begin{widetext}
\begin{eqnarray}
\mathcal{L}_{ew}&=&\frac{1}{\sqrt{\Lambda}}\left\{\lambda_1[E_1^c C_5^{-1}L_1\phi^\dagger +U_1^c
C_5^{-1}Q_1\tilde{\phi}^{\dagger}]+\lambda_2[E_2^c C_5^{-1}L_2\phi^\dagger
+U_2^c C_5^{-1}Q_2\tilde{\phi}^{\dagger}]+\right.\nonumber\\
& &\left.\lambda_1'[N_1^c C_5^{-1}L_1\tilde{\phi}^{\dagger}+D_1^c C_5^{-1}Q_1\phi^{\dagger}]+\lambda_2'[N_2^c C_5^{-1}L_2\tilde{\phi}^{\dagger} +D_2^c C_5^{-1}Q_2\phi^{\dagger}]+\mathrm{H.c.}\right\},\label{bulk_5d_ql_ew_yukawa_Lagrangian}
\end{eqnarray} 
\end{widetext}
where the $\lambda$'s are flavour space Yukawa coupling matrices.
This Lagrangian generates fermion Dirac mass terms when $\phi^{(0)}$
develops a VEV, with the SM fermions acquiring mass through the
$\lambda_1$, $\lambda_1'$ terms. If the SM fermions have uniform
profiles along the extra dimension, then the troublesome tree level mass
relations $m_e=m_u$ and $m_\nu=m_d$ arise. However, if one is able to
generate different five dimensional wavefunction profiles for quarks
and leptons, the troublesome tree level relations will not persist in
the effective four dimensional theory. We shall have more to say on
this matter in Section~\ref{sec:bulk_5d_ql_fermion_mass}, but it will prove useful to consider the
stability of the proton within our model first.
\section{Proton Decay\label{sec:bulk_5d_ql_p_decay}}
At the five dimensional level, the lowest dimension non-renormalizable
operator which leads to proton
decay in our model is
\begin{eqnarray}
\frac{f}{\Lambda^{9/2}}\epsilon_{\alpha\beta\gamma}Q_2^\alpha Q_2^\beta Q_2^\gamma \chi^{\dagger}_{\bar{\alpha}}L_1^{\bar{\alpha}}.\label{orb_ql_bulk_p_decay_operator_5d}
\end{eqnarray}
where $\alpha$, $\beta$, $\gamma$ are quark colour indices,
$\bar{\alpha}$ is the lepton colour index and $f$ is a dimensionless
coupling constant. If the SM quarks and
leptons have uniform profiles across the extra dimension, proton
decay occurs in the effective four dimensional theory via
the operator
\begin{eqnarray}
\frac{f}{(\Lambda \pi R)^{3/2}}\frac{w}{\Lambda^3}\epsilon_{\alpha\beta\gamma}Q^\alpha Q^\beta Q^\gamma L,\label{orb_ql_bulk_p_decay_operator_5d_2}
\end{eqnarray}
where $Q$ and $L$ denote four dimensional quark and lepton fields
respectively.  Current experimental bounds
require the lifetime of the proton to be in excess of
$1.6\times10^{33}$ years~\cite{Nath:2006ut}. With $w\sim1$~TeV and
$\Lambda R\sim100$ one requires $\Lambda\sim 5\times 10^{10}$~GeV to suppress
the proton decay rate.

It is possible to reduce the scale of the cut-off by
localising quarks and leptons at different points in the extra
dimension~\cite{Arkani-Hamed:1999dc}. If quarks and leptons are
separated in the extra dimension, the diminished overlap of the quark
and lepton wavefunctions in the fifth dimension serves to reduce the
proton decay rate. To suppress the proton decay
rate we shall localise quarks and leptons at different fixed
points. To illustrate our idea it will suffice to consider an
$S^1/Z_2$ orbifold. Let us add a
gauge singlet, bulk scalar field $\Sigma$ which transforms as
\begin{eqnarray}
\Sigma(x^\mu,y)&\rightarrow& \Sigma(x^\mu,-y)=-\Sigma(x^\mu,y),\label{bulk_ql_sigma_bcs}
\end{eqnarray}
under the $Z_2$ symmetry and has the potential
\begin{eqnarray}
V(\Sigma)=\frac{\kappa}{4\Lambda}(\Sigma^2-v'^2)^2,
\end{eqnarray}
where $\kappa$ ($v'$) is a dimensionless (dimensionful) constant. The
minimum of the potential clashes with the orbifold boundary conditions
(\ref{bulk_ql_sigma_bcs})
and results in the vacuum profile~\cite{Georgi:2000wb} (see also~\cite{Perez:2002wb}),
\begin{eqnarray}
\langle\Sigma\rangle(y)\approx \frac{v}{\sqrt{2\pi R}}\tanh[\xi(-\pi R-y)]\tanh[\xi
y],
\end{eqnarray}
where $\xi^2=\kappa v^2/2$ and we have introduced $v=\sqrt{2\pi R}v'$. The points $y=\pm \pi R$ are
identified under the
orbifold construction. One can simplify the analysis when $\kappa v^2(2\pi
R)^2\gg1$ by treating the VEV profile of $\Sigma$ as a step
function~\cite{Kaplan:2001ga},
\begin{eqnarray}
\langle\Sigma\rangle(y)=\frac{v}{\sqrt{2\pi R}} h(y),
\end{eqnarray}
where,
\begin{eqnarray}
h(y)=\left\{\begin{array}{cc} +1& \pi R>y>0 \\-1& -\pi R<y<0.
\end{array}\right.
\end{eqnarray}
In five dimensions, gauge
invariance permits the Yukawa couplings
\begin{eqnarray} 
-\frac{h_F}{\sqrt{\Lambda}} \bar{F}F\Sigma,
\end{eqnarray}
for all fermion fields,  where $h_F$ is a Yukawa
coupling constant. The shape of zero mode fermions in the extra dimension, when
$h_Fv>0$, is subsequently given by~\cite{Kaplan:2001ga}
\begin{eqnarray}
 F^{(0)}_L=\sqrt{\frac{2|h_F v|}{1-e^{-2|h_F v|\pi R}}}e^{-|h_F v|y},\label{orb_ql_bulk_zero_mode_fermi_profile}
\end{eqnarray}
or via the replacement $y\rightarrow (\pi R-y)$ if \mbox{$h_F v<0$},
demonstrating the localisation of zero mode fermions at different
fixed points on the $S^1/Z_2$ orbifold, depending on the sign of the
coupling constant $h_F$.

We wish to suppress the proton decay rate by localising quarks and
leptons at different fixed points. Let us take $\Sigma$ to be odd
under the discrete QL symmetry~\cite{Coulthurst:2006bc}. The resulting Yukawa Lagrangian is,
\begin{eqnarray}
\mathcal{L}_{\Sigma}&=&\frac{1}{\sqrt{\Lambda}}\sum_{i=1,2}\left\{h_{D_i}[D^{c2}_i -
  N^{c2}_{i}] +h_{U_i}[U^{c2}_i -
  E^{c2}_{i}] \right.\nonumber\\
& &\left.\mkern25mu + h_{Q_i}[Q^{2}_i -
  L^{2}_{i}]\right\}\Sigma,\label{orb_ql_bulk_localizer_lagrangian_p_decay}
\end{eqnarray}
where $h_{U_i}$, $h_{D_i}$ and $h_{Q_i}$ are Yukawa coupling
constants. We use an obvious notation with $F^2=\bar{F}F$ and we have
suppressed family indices. If one takes all
Yukawa couplings to be greater than zero, quarks and leptons are
automatically localised at different fixed points. 

After integrating
over the extra dimension, the operator
(\ref{orb_ql_bulk_p_decay_operator_5d}) produces a proton
decay inducing operator in the four dimensional theory. The
approximate form of this operator is
\begin{eqnarray}
\frac{f'}{\sqrt{\Lambda R}}\frac{wv}{\Lambda^2}\exp\left\{-cv\pi R\right\}
\frac{Q^3L}{\Lambda^2},
\end{eqnarray}
where $f'$ and $c$ are dimensionless constants and $Q$ and $L$ are
four dimensional
quark and lepton fields respectively. For $\Lambda$ of order $100-500$~TeV and an order TeV QL symmetry breaking scale,
the lower bound on the lifetime of the proton requires $v\pi R>40$ if one takes
$c$ to be order unity.
\section{Fermion Mass\label{sec:bulk_5d_ql_fermion_mass}}
We have seen that one is able to understand
the long lifetime of the proton if quarks and
leptons are localised at different fixed points. In
Section~\ref{sec:bulk_5d_ql_yukawas} we noted that despite the Yukawa
coupling relations induced by the QL
symmetry in equation (\ref{bulk_5d_ql_ew_yukawa_Lagrangian}), one may 
understand the absence of relations of the type $m_e=m_u$ in the
effective four dimensional theory if quarks and leptons have different
wavefunction profiles in the extra dimension.

Inspection of equations
(\ref{orb_ql_bulk_zero_mode_fermi_profile}) and
(\ref{orb_ql_bulk_localizer_lagrangian_p_decay}) reveals that the
methods employed to separate quarks and leptons in the extra dimension
induce identical fifth dimensional wavefunction profiles for fermions
related by the QL symmetry. Upon integrating over the
fifth dimension the troublesome tree level mass relationships persist,
despite quarks and leptons being localised at different fixed
points. We could in principle add a second SM Higgs doublet to the
model and generate enough parameter freedom to remove the unwanted
mass relations~\cite{Foot:1990dw}. However this will not allow us to understand the
lightness of the known neutrinos relative to the electroweak scale.

In this
section we shall apply a purely higher dimensional mechanism to remove the
undesirable tree level
mass relations. The
full construction of a theory of flavour is beyond the scope of the
present work. We shall sketch our ideas in what follows and to this
end it is helpful to recall some key results obtained in previous
studies involving split fermions.

The work of Arkani-Hamed and
Schmaltz (AS)~\cite{Arkani-Hamed:1999dc} demonstrated that four dimensional
flavour could be
addressed in terms of fifth dimensional wavefunction overlaps. AS
spatially separated the left- and right-chiral fermion fields in an
extra dimension, with the size of the spatial separation influencing
the degree of
wavefunction overlap. The amount of overlap then determined the size
of the effective four dimensional Yukawa couplings to the SM
Higgs. However one need not separate the left- and right-chiral fields
to address flavour in terms of fifth dimensional wavefunction
overlaps. Indeed one can localise all the fermions
at one point in the extra dimension, provided the left- and
right-chiral fermions have fifth dimensional wavefunctions with
different widths~\cite{Lillie:2003sq}.

We shall employ each of these mechanisms to
realize flavour. In particular we shall localise all quark fields at
one point in the extra dimension, with the different widths of the
left- and right-chiral quark fields
determining the effective four dimensional Yukawa couplings in the
quark sector. For the leptons we shall separate the left- and
right-chiral fields in the extra dimension, thus motivating the
observed flavour differences between the quark and lepton sectors.

AS separated fermion fields by introducing distinct bulk Dirac
mass terms for the different bulk fermions. In an orbifold theory,
however, the fermion transformations necessary to introduce four
dimensional chirality preclude bulk Dirac mass terms. One may localise
fermions at non-fixed points in an orbifold theory by introducing a
second localising scalar~\cite{Grossman:2002pb}. This works as
follows~\cite{Grossman:2003sr}. With one localising scalar the chiral
zero mode of a fermion $F$
is localised at one of the orbifold fixed points. The point of
localisation is determined by the sign of the product $h_F v$ as
mentioned already
in Section~\ref{sec:bulk_5d_ql_p_decay}, which amounts to the sign of
$h_F$ when $v>0$. If the field $F$ couples to a
second bulk scalar with an opposite sign Yukawa coupling, the second
scalar tends to localise the zero mode at the opposite fixed
point. When one scalar is dominant the fermion is found localised at
the fixed point preferred by that scalar. In general, however, an interplay
between the two scalars results in a compromise which sees the fermion
localised in the bulk. More technical details may be found
in~\cite{Grossman:2002pb}.

Let us now investigate this idea in the QL symmetric framework,
assuming an $S^1/Z_2$ orbifold again for simplicity. We
introduce a second bulk scalar, $\sigma$, which transforms as
\begin{eqnarray}
\sigma(x^\mu,y)&\rightarrow& \sigma(x^\mu,-y)=-\sigma(x^\mu,y),
\end{eqnarray}
under the orbifold $Z_2$ symmetry and transforms trivially under the
QL symmetry. The Yukawa Lagrangian for $\sigma$ is,
\begin{eqnarray}
\mathcal{L}_{\Sigma}&=&\frac{1}{\sqrt{\Lambda}}\sum_{i=1,2}\left\{f_{D_i}[D^{c2}_i +
  N^{c2}_{i}] +f_{U_i}[U^{c2}_i +
  E^{c2}_{i}] \right.\nonumber\\
& &\left.\mkern25mu + f_{Q_i}[Q^{2}_i +
  L^{2}_{i}]\right\}\sigma.\label{orb_ql_bulk_localizer2_lagrangian_p_decay}
\end{eqnarray}
We achieved quark-lepton separation in the last section by demanding
\begin{eqnarray}
h_{U_i},h_{D_i}, h_{Q_i}>0.
\end{eqnarray}
Let us further demand that
\begin{eqnarray}
f_{U_i},f_{D_i}, f_{Q_i}>0,
\end{eqnarray}
so that all quark Yukawa couplings to $\Sigma$ and $\sigma$ are
positive.

Consider first the effects of the second scalar $\sigma$
on the quark sector. Inspection of
(\ref{orb_ql_bulk_localizer_lagrangian_p_decay}) and
(\ref{orb_ql_bulk_localizer2_lagrangian_p_decay}) reveals that both
$\Sigma$ and $\sigma$ will attempt to localise the quarks at the same
fixed point. Thus for all positive values of the Yukawa couplings, which we
generically denote as $h_F$ and $f_F$, the quark fields will be
localised at one fixed point. Note that the SM fields $Q$, $u_R$ and
$d_R$ will each have, in general, different profiles in the extra
dimension, allowing one to reproduce the necessary quark flavour
spectrum along the lines of~\cite{Lillie:2003sq}.

Let us now consider the lepton sector. This sector is more complicated
because $\Sigma$ and $\sigma$ attempt to localise a given chiral zero
mode lepton at different fixed
points. We have shown in the last section that the long lifetime of
the proton may be understood if quarks and leptons are localised at
different fixed points. To retain this result we shall require
$\Sigma$ to dominate over $\sigma$ for the SM fields $L$ and
$e_R$. The situation with $\nu_R$ is not as clear. We shall discuss
neutrino masses further in the next section, but for now let us estimate
the degree of
separation required between the field $L$ and the field $e_R$ to
obtain an order MeV electron mass. It was shown
in~\cite{Grossman:2002pb} that, in the decoupling limit
of the two localising scalars,  one may arrange zero mode
fermions to be localised in the bulk of an orbifold theory with fifth
dimensional profiles of the form
\begin{eqnarray}
G^{(0)}(y)=N\exp\left\{-\frac{k}{2}\frac{v^2}{\Lambda} (y-y_m)^2\right\}.\label{bulk_ql_perez_gaussian}
\end{eqnarray} 
Here $N$ is a normalization factor, $v$ denotes the VEV of one of
the localising scalars and $y_m$ is the location of the wavefunction
maxima. We show in Section~\ref{sec:bulk_ql_bounding_scales} that we
expect $v\ge 420$~TeV and use this lower bound in what
follows. To leading order the dimensionless constant $k$ depends on the
largest fermion-bulk scalar
Yukawa coupling and the dimensionless scalar quartic
self-coupling $k\approx h_F \sqrt{2 \kappa}$. Note that naive
dimensional analysis permits values of $k\sim 24\pi^3\sqrt{2}\sim
10^{3}$ if the
underlying theory has strong couplings at the
cut-off~\cite{Kaplan:2001ga,Chacko:1999hg}. However we shall be able
to obtain acceptable suppression of the electroweak scale with
significantly lower $k$ values~\cite{Kaplan:2001ga}.

The charged lepton mass terms arise
from the couplings 
\begin{eqnarray}
\frac{\lambda_1}{\sqrt{\Lambda}}E_1^c C_5^{-1}L_1\phi^\dagger,
\end{eqnarray}
which lead to Dirac masses of the form
\begin{eqnarray}
\frac{\lambda_1 u}{\sqrt{\Lambda \pi R}}\int_{0}^{\pi
  R}e_R^{(0)}(y)L^{(0)}(y) dy,\label{bulk_5d_ql_split_dirac_mass}
\end{eqnarray}
where
$e_R^{(0)}(y)$ is the electron wavefunction in the extra dimension
which takes the form of $G^{(0)}(y)$ in Equation~(\ref{bulk_ql_perez_gaussian}). The localisation of the zero mode of the field $L$ is assumed to
be dominated by the scalar $\Sigma$ such that $L^{(0)}(y)$ has the
exponential form of equation
(\ref{orb_ql_bulk_zero_mode_fermi_profile}).

Taking $\lambda,k\sim1$ in (\ref{bulk_5d_ql_split_dirac_mass}) reveals that
a charged lepton
Dirac mass of order MeV is obtained if $e_R$ is localised a
distance of $\pi R/15$ from the fixed point at which $L$ is
localised. Thus one is not required to shift the zero modes
corresponding to the SM right-chiral charged leptons far to obtain
a realistic charged lepton spectrum. This slight shift from the fixed
point should not produce an
observable enhancement of the proton decay rate.
\section{Neutrino Masses\label{bulk_ql_nu_mass}}
Neutrinos acquire Dirac masses at the electroweak symmetry
breaking scale via the coupling
\begin{eqnarray}
\frac{\lambda_1'}{\sqrt{\Lambda}}N_1^c
C_5^{-1}L_1\tilde{\phi}^{\dagger}.
\end{eqnarray}
We wish to suppress the electroweak contribution to the neutrino mass
by separating $\nu_R$ and $\nu_L$ in the extra dimension. Given that
neutrino masses are required to be less than an eV it is clear
that one must localise the SM neutrinos away from the quarks to
suppress the proton decay mode
$p\rightarrow\nu_L\pi$. We shall shift $\nu_R$ into the bulk to
suppress the electroweak contribution to the neutrino mass, however
the acceptable proximity of $\nu_R$ to the quarks is not as clear. We
must consider the mass scale of this field to determine its
acceptable proximity to the quarks, and this requires a consideration
of the non-renormalizable contributions to the neutrino mass sector.

Consider the non-renormalizable operators which
induce Majorana mass for the neutrinos in our model.  
The operator
\begin{eqnarray}
\mathcal{O}_{\nu_L}= \frac{g}{\Lambda^5}(\chi^\dagger
L_1\tilde{\phi}^{\dagger})^2,\label{bulk_5d_ql_nuetrino_mass_operator}
\end{eqnarray}
written in terms of five dimensional fields, with $g$ a dimensionless
constant, has
dimension ten and in the effective four
dimensional theory induces the operator
\begin{eqnarray}
\mathcal{O}_{\nu_L}^{eff}= \frac{g}{(\Lambda \pi R)^2}\frac{(\chi^{(0)\dagger}
L^{(0)}\tilde{\phi}^{\dagger(0)})^2}{\Lambda^3},\label{bulk_5d_ql_eff_nuetrino_mass_operator}
\end{eqnarray}
which has dimension seven. When
$\chi^{(0)}$ and $\phi^{(0)}$ develop VEV's,
$\mathcal{O}_{\nu_L}^{eff}$ produces Majorana masses for the SM
neutrinos $\nu_L$. These masses take the form
\begin{eqnarray}
m_{\nu_L}=\frac{g}{(\Lambda \pi R)^2}\frac{(wu)^2}{\Lambda^3}\label{bulk_ql_sm_neutrin_mass}.
\end{eqnarray}
We require $m_{\nu_L}\sim g\times 1$~eV so that this mass is
in the right range to accommodate
the atmospheric and solar neutrino oscillation data. The
neutrinos $\nu_R$ also
acquire mass at the
non-renormalizable level. At the five dimensional level, one has the operator
\begin{eqnarray}
\mathcal{O}_{\nu_R}=\frac{g'}{\Lambda^2}(\chi N^c_1)^2,
\end{eqnarray}
where $g'$ is a dimensionless coupling constant. This produces a
Majorana mass for $\nu_R$ in the effective four dimensional theory,
\begin{eqnarray}
m_{\nu_R}=\frac{g'}{(\Lambda
  \pi R)}\frac{w^2}{\Lambda}\label{bulk_ql_sterile_nu_mass}.
\end{eqnarray}
We intend to reduce the effective Dirac mass between $\nu_L$ and
$\nu_R$ in the four dimensional theory below the electroweak scale by
separating $L$ and $\nu_R$ in the extra dimension. To estimate the degree of suppression
required of the effective Dirac mass we shall take $m_{\nu_R}$ to be
  $O(\mathrm{GeV})$ (we shall see in
  Section~\ref{sec:bulk_5d_ql_sterile_cosmology} that masses in this
  range are compatible with our framework). Consider the complete neutrino mass matrix in the
Majorana basis for one generation,
\begin{eqnarray}
\left(\begin{array}{cc}m_{\nu_L}&m_D\\
      m_D&m_{\nu_R}\end{array}\right),
\end{eqnarray}
where $m_D$ is the effective four dimensional Dirac
mass. Under the usual see-saw hierarchy
\begin{eqnarray}
m_{\nu_L}\ll m_D\ll m_{\nu_R},
\end{eqnarray}
the mass eigenvalues take the approximate form
\begin{eqnarray}
m_1&\approx& m_{\nu_L}-m_D^2/m_{\nu_R},\\
m_2&\approx& m_{\nu_R}.
\end{eqnarray}
The seesaw contribution to the light neutrino mass eigenvalue,
$m_D^2/m_{\nu_R}$, is required to be order eV or less to ensure that
$m_1$ is not too heavy. Hence one requires $m_D\lesssim1$~keV
for $m_{\nu_R}\sim1$~GeV to
ensure that $m_D^2/m_{\nu_R}\lesssim10^{-1}$~eV. This requires $\nu_R$ to be localised a distance $\gtrsim\pi R/12$ from
the boundary at which $L$ is localised. We show in
Section~\ref{sec:bulk_5d_ql_sterile_cosmology} that $m_{\nu_R}$ is
  required to be at least of order GeV. Thus proton decays
  into $\nu_R$ are kinematically forbidden and localising $\nu_R$
  further into the bulk does not effect the stability of the proton, a
  result which holds even if $\nu_R$ is localised at the `quark end' of
  the extra dimension. 
\section{The Size of the Extra
  Dimension\label{sec:rev_bulk_ql_bounding_r}}
We now consider the bounds on $R$, the
compactification scale. In Section~\ref{sec:bulk_5d_ql_gauge_sector}
it was shown that the mass of the lightest $Y$ boson has a
contribution inversely proportional to $R$ (the usual KK contribution). In 4D~\cite{Foot:1991fk} and 5D~\cite{McDonald:2006dy}
QL models explored previously, bounds on
the $Y$ boson mass have been obtained by considering the decay
$\mu\rightarrow 3e$, which proceeds radiatively
by the creation of virtual $Y$ bosons. This results in an approximate
bound of $M_Y\ge 5$~TeV. In~\cite{McDonald:2006dy} this led to a
bound on $R$, however the current
model requires a re-evaluation of this bound.

We note from
Section~\ref{sec:bulk_5d_ql_fermion_parities} that the liptons which
form $SU_l(3)$ multiplets with SM leptons do not possess zero
modes. Vertices in the effective 4D Lagrangian which couple SM leptons
to liptons via $Y$ boson exchange will generally take the form
\begin{eqnarray}
\mathcal{L}\sim K_{l L'}YlL',
\end{eqnarray}
where $Y$, $L'$ and $l$ denote $Y$ boson, lipton and SM lepton
operators respectively and $K_{l L'}$ is a numerical factor
representing the wavefunction overlap of $Y$, $l$ and $L'$ in the
extra dimension. The lightest lipton that couples to a SM lepton $l$
via $Y$ exchange is one of the $n=1$ KK liptons in the
same $SU_l(3)$ multiplet as $l$. However the bulk scalars employed to
localise chiral zero mode fermions in
Sections~\ref{sec:bulk_5d_ql_p_decay}
and~\ref{sec:bulk_5d_ql_fermion_mass} also alter the mass and 5D
profiles of $n>0$ KK modes. When one localising scalar is employed it
has been shown that the low lying KK modes (with
$n\ne 0$) also become localised, albeit with broader 5D
wavefunctions~\cite{Hung:2003cj}. Furthermore the odd and even
KK modes tend to be localised at opposite fixed points on an
$S^1/Z_2$ model. Given that we localise SM leptons at
one boundary of the extra dimension, the $n=1$ KK liptons will be
found at the other boundary. Thus $K_{l L'}\sim \exp\{-v\pi R\}$ for
$n=1$ liptons, rendering the associated vertex vanishingly
small. Consequently the
radiative decay $\mu\rightarrow 3e$ will occur only via the
coupling of SM charged leptons to even $n$ KK mode liptons, with the dominant
contribution arising from the $n=2$ mode.

It is difficult to evaluate $K_{l L'}$ exactly for an $n=2$ lipton
$L'$ in our model, given the two localising scalars employed. This
would require a determination of the fermion 5D wavefunction profiles for a
two localising scalar scenario, which is beyond the scope of the
present study. To
approximate $K_{l L'}$ we use the one bulk scalar results
of~\cite{Hung:2003cj} where analytic expressions for the KK tower 5D
wavefunctions were obtained. We find that the overlap is
very small for a range of parameter values, with $K_{l L'}\sim10^{-3}$
for a zero mode lepton $l$ and
an $n=2$ lipton $L'$. Assuming similar values for our model
we find that the decay $\mu\rightarrow 3e$ is highly suppressed due to
the factor of $K_{l L'}^4$ in the rate. The associated bound on $M_Y^{(1)}$, and
thus $R$, is exceptionally weak.

A stronger bound on $R$ may be obtained by considering Flavour
Changing Neutral Currents (FCNCs). It is a generic feature of models
involving split fermions that FCNCs arise~\cite{Delgado:1999sv}. Zero mode gauge bosons
possess uniform profiles in the extra dimension and thus couple
uniformly to localised fermions. However the $n>0$ KK mode gauge
bosons are not uniform along the extra dimension and thus, in
general, couple to different localised fermions with different
strengths, resulting in FCNCs.

The most stringent bounds from FCNCs in
split fermion models arise in the quark sector, where exchange of KK
mode gluons can lead to bounds of
$1/R>5000$~TeV~\cite{Delgado:1999sv}. We have localised all quarks at
one fixed point, aiming to realize quark flavour through varying width
5D quark wavefunctions. This significantly reduces the bounds from
FCNCs in the quark sector to $1/R\ge
2-5$~TeV~\cite{Lillie:2003sq}. In the lepton sector we have localised
all left-chiral SM leptons at one fixed point, whilst the SM right-chiral
leptons are localised near the fixed point, with different points of
localisation for different fields. Thus only gauge bosons which couple
to right-chiral fields, namely the hypercharge and $U_{X'}(1)\subset
SU_l(3)$ gauge bosons, will have markedly different couplings to
different leptons and thus give rise to FCNCs in the lepton sector.

By considering the contribution of the KK tower for the electrically neutral
$SU_L(2)$ gauge boson to
$\mu\rightarrow 3e$ a bound of $1/R\gtrsim 30$~TeV has previously been
obtained~\cite{Delgado:1999sv}. In our model the KK tower for the $Z'$
gauge bosons will
contribute to $\mu\rightarrow 3e$ and we expect this bound to apply,
even though KK $SU_L(2)$ gauge bosons
will not mediate this decay. Thus we require $1/R\gtrsim 30$~TeV to
avoid trouble with FCNCs in the model. We note that if all right-chiral
SM leptons
were localised at one point the lower bound of $1/R\ge
2-5$~TeV would apply, though this would have to be arbitrarily imposed on
the model.
\section{Cosmology of the sterile
  neutrino\label{sec:bulk_5d_ql_sterile_cosmology}}
We have seen in Section~\ref{sec:bulk_5d_ql_fermion_mass} that the
neutrinos will acquire Majorana masses at the non-renormalizable
level. The cosmological density of the neutrino $\nu_R$ must be
considered to ensure that this state does not spoil any of the
successful features of the standard cosmological model. The neutrinos
$\nu_R$ will remain in thermal equilibrium in the early universe. This
results from annihilation's involving $Z'$ bosons, namely:
\begin{eqnarray}
\nu_R\bar{\nu}_R \leftrightarrow Z' \leftrightarrow l \bar{l},
\end{eqnarray}
where $l$ denotes a SM lepton. At some temperature the state $\nu_R$
will freeze out and the remaining cosmological abundance will
contribute to the energy density of the universe. The perturbations to
the usual scenario induced by $\nu_R$ depend on its mass and
lifetime. It is known from 4D models that right-chiral neutrinos may
decay via photon emission or via $SU_l(2)\subset SU_l(3)$ glueball
emission.

Before we enter into the specifics, let us summarise the
findings of this section for readers not interested in the more
technical details. Below we show that in our 5D model the neutrino
decay $\nu_R\rightarrow\nu_L\gamma$ is expected to dominate the
decays involving glueball emission. Known cosmological bounds
between the mass of heavy neutrinos and the lifetime for radiative
decays then require $m_{\nu_R}\ge100$~GeV. This result is obtained
under the assumption that the neutrino Dirac mass $m_D$ is suppressed
to keV energies
by localising $\nu_R$ in the bulk. However the lightest $\nu_R$
effectively becomes
stable if $m_D$ is suppressed by
localising $\nu_R$ at the quark end of the extra dimension. In this
case the usual bound for stable massive neutrinos applies and one
requires $m_{\nu_R}\gtrsim5$~GeV.  Let us now consider these decay
mechanisms in more detail. 
\subsection{Radiative Heavy Neutrino Decays\label{bulk_ql_radiative_nu_decay}}
\begin{figure}[b] 
\centering
\includegraphics[width=0.33\textwidth]{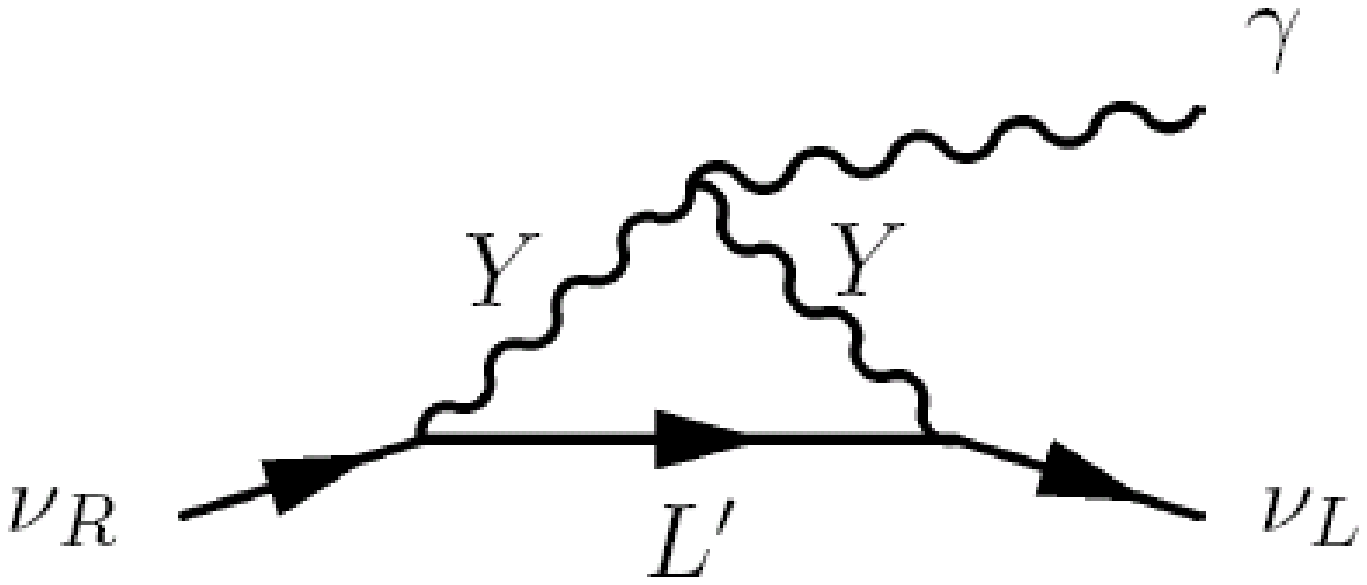} 
\includegraphics[width=0.33\textwidth]{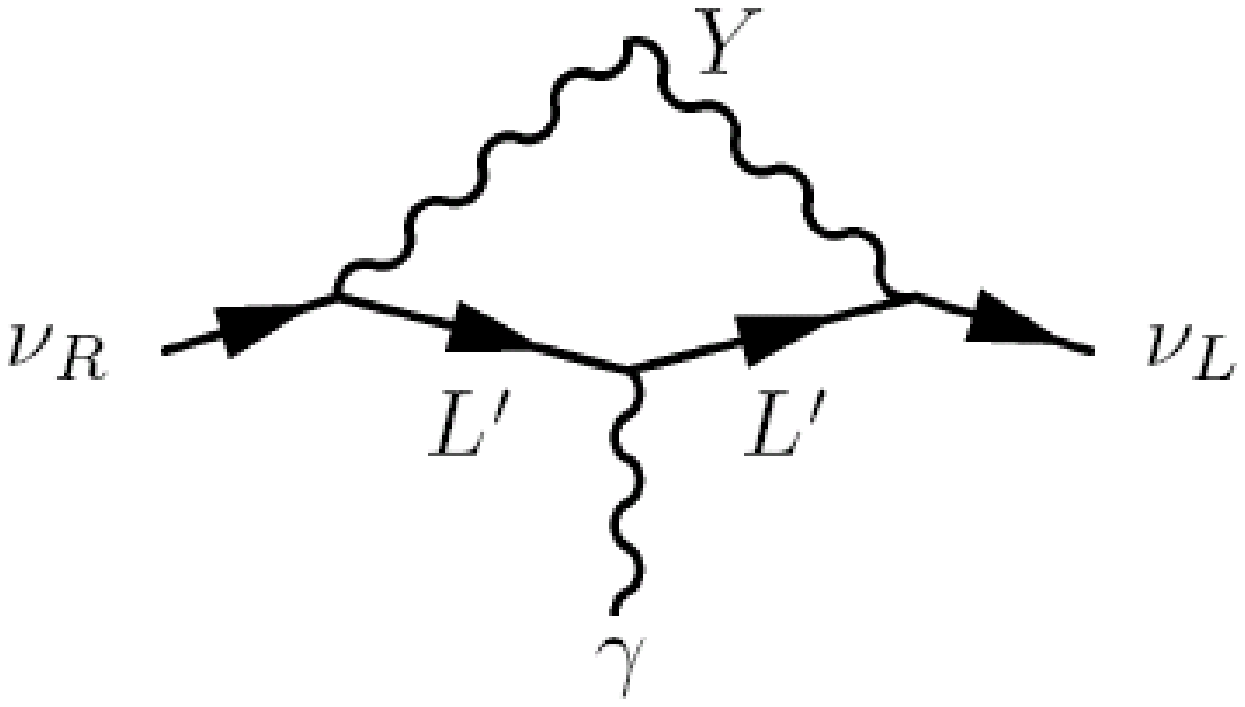}
  \caption{The graphs for $\nu_R\rightarrow\nu_L+\gamma$ in
    quark-lepton symmetric models.}
  \label{fig:neutrinodecay}
\end{figure}

Consider the decay $\nu_R\rightarrow \nu_L\gamma$, where $\nu_L$
denotes a SM neutrino. In QL models this occurs radiatively via the
creation of a $Y$ boson and a lipton (see Figure~\ref{fig:neutrinodecay}).
The rate for
these decays can be obtained by modifying existing results for
similar neutrino decays. In particular
reference~\cite{Goldman:1977jx} studied radiative neutrino decays of
the type $\nu_{\mu}\rightarrow \nu_{e}
\gamma$ when exotic leptons and gauge bosons exist. Equation (2) of
that work identifies the rate for $\nu_{\mu}\rightarrow \nu_{e}
\gamma$ to be
\begin{eqnarray}
\Gamma(\nu_\mu\rightarrow\gamma \nu_e)= \frac{G_F^2 \alpha B^2
  M_L^2M_{\nu_\mu}^3}{128 \pi^4}\label{goldman_bound},
\end{eqnarray}
where $G_F$ is the Fermi constant, $\alpha$ is the fine structure
constant, $B$ is a numerical vertex factor and $M_L$ ($M_{\nu_\mu}$) is
the mass of an exotic lepton (muon neutrino). This expression may be
applied to the decay $\nu_R\rightarrow \gamma \nu_L$ in the 5D QL model
by making the following replacements
\begin{eqnarray}
\alpha&\rightarrow& \frac{1}{4}\alpha,\label{goldman_replacement_alpha}\\
M_{\nu_\mu}&\rightarrow& m_{\nu_R},\\
B^2G_F^2&\rightarrow& K_1^2K_2^2G_Y^2\sin^2\theta (\sqrt{2})^4.\label{goldman_replacement_K_1_2_G_y}
\end{eqnarray}
Here (\ref{goldman_replacement_alpha}) is required as the $Y$ bosons
and liptons
carry electric charge $1/2$. The factor of $\sqrt{2}$ is due to the
enhanced coupling of the $n>0$ KK mode $Y$ boson and $G_Y$ is the
equivalent of the Fermi constant for the $Y$ bosons,
\begin{eqnarray}
\left(\frac{G_Y}{\sqrt{2}}\right)^2 =\left(\frac{g_S^2}{8
    M_Y^{2(1)}}\right)^2.
\end{eqnarray}
The angle $\sin \theta$ represents the mixing between a lipton which
couples to $\nu_R$ and a lipton which couples to $\nu_L$. This mixing
is induced by electroweak symmetry breaking and vanishes in the limit
$u\rightarrow 0$. The angle takes the approximate form
\begin{eqnarray}
\sin\theta\approx \frac{K_3u}{M_L},
\end{eqnarray}
where $K_3$ is the 5D wavefunction overlap of the $\nu_R$ lipton with
the $\nu_L$ lipton. We shall comment further on $K_3$ shortly. The low
lying $n>0$ KK mode liptons will have mass of order $v$, the bulk scalar
symmetry breaking scale~\cite{Hung:2003cj}. Note that the interplay of
the $M_L^2$ factor in the numerator of (\ref{goldman_bound}) and the
$M_L^{-1}$ dependence of $\theta$ renders the rate for
$\nu_R\rightarrow \gamma \nu_L$ independent of $v$. We expect the $n=2$ KK
lipton to be the
lightest state to contribute significantly to the decay
$\nu_R\rightarrow \gamma \nu_L$. The $n=1$ state is not localised at
the same fixed point as the $n=0$ chiral modes~\cite{Hung:2003cj} and thus its
contribution to the decay will be suppressed by the $K_3$ factor.

The factors $K_1$ and $K_2$ in equation
(\ref{goldman_replacement_K_1_2_G_y}) represent the
wave function overlap between
$n=0$ chiral mode neutrinos, the $n=2$ lipton which couples to
this neutrino and the $n=1$ $Y$ gauge boson. They take the form
\begin{eqnarray}
K_{1}&\sim&\int_0^{\pi R/2}\nu_R^{(0)}(y)L'^{(2)}(y)\cos
\frac{y}{R}dy,\nonumber\\
K_{2}&\sim&\int_0^{\pi R/2}L^{(0)}(y)L'^{(2)}(y)\cos\frac{y}{R}dy,
\end{eqnarray}
where $\nu_R^{(0)}(y)$ and $L^{(0)}(y)$ are zero mode SM singlet neutrino and
electroweak lepton doublet 5D wavefunctions respectively. $L'^{(2)}$ represents an
$n=2$ lipton 5D wavefunction and the $Y$ boson fifth dimensional
profile is given by the cosine factor. The absence of
analytic expressions for the 5D fermion KK tower profiles in two bulk
scalar models and the dependence on free parameters makes it difficult
to determine these factors
exactly. We approximate these using the results obtained for the one
localising scalar case in~\cite{Hung:2003cj} and
find that typical values are $K_{1,2}\sim10^{-3}$.

To estimate $K_3$ we consider the case where $\nu_R$ and $\nu_L$ are
localised by one bulk scalar, but the potential trapping $\nu_R$ in
the bulk has a minimum shifted away from the fixed point. Noting
that $\nu_R$ is shifted some distance away from the fixed point such
that the effective Dirac mass coupling $\nu_R$ and $\nu_L$ in the 4D
theory is $m_D$ and using the analytic $n=2$ KK mode wavefunctions
of~\cite{Hung:2003cj} we find that typically $K_3 u \sim 10^2
m_D$. The Dirac mass coupling the $n=2$ liptons is expected to be
larger than the associated $n=0$ mode Dirac mass $m_D$ as the 5D
wavefunctions for the higher KK
modes generally have broader wavefunctions in the extra dimension than the
zero modes.

Putting all this together we find that
\begin{eqnarray}
\tau(\nu_R\rightarrow \gamma \nu_L)=5\times 10^{-11}(s\ GeV)
\frac{M_Y^{4(1)}}{m_D^2m_{\nu_R}^3}.\label{bulk_ql_radiative_nu_r_lifetime}
\end{eqnarray}
Taking $m_D\sim 1$~keV and $1/R\sim 30$~TeV gives $\tau\sim10^{22}s$
($\tau\sim 10^{19}s$) for $m_{\nu_R}=10^2$MeV (1~GeV). Such long lived
neutrinos would still be decaying and would contribute to the diffuse
photon background. Studies of the diffuse photon background require
heavy neutrinos to satisfy the bound
\begin{eqnarray}
\tau\ge (10^{25}s\ GeV^2)\times m^{-2}
\end{eqnarray}
if their lifetime exceeds $\tau\ge 3\times
10^{17}s$~\cite{Kolb:1990vq}. Thus one requires
$\tau\ge 10^{27}s$ ($10^{25}s$) for $m_{\nu_R}=10^2$MeV (1~GeV), which is
not satisfied by the radiative $\nu_R$ decays. Note that increasing
$m_{\nu_R}$ decreases the radiative lifetime of $\nu_R$. For
$m_{\nu_R}\ge 10$~GeV one finds that $\tau\le10^{17}s$. If the
lifetime of a heavy neutrino lies in the range $t_\mathrm{rec}\le \tau\le t_U$, where $t_{\mathrm{rec}}$ ($t_U$)
  is the time of recombination (age of the universe), a different
  relationship between the neutrinos mass and lifetime must be satisfied. This
  relationship takes the form~\cite{Kolb:1990vq} 
\begin{eqnarray}
m_{\nu_R}\gtrsim 8\times 10^{-3}\tau^{1/3}_{\mathrm{sec}}~\mathrm{GeV}.
\end{eqnarray}
Using
  (\ref{bulk_ql_radiative_nu_r_lifetime}) gives
\begin{eqnarray}
m_{\nu_R}\gtrsim (8\times 10^{-3})^{1/2}\left(\frac{5\times 10^{-11}
    M^{4(1)}_Y}{m_D^2}\right)^{1/6},\label{heavy_nu_bound_bulk_ql_2}
\end{eqnarray}
where all masses should be given in GeV. Using the lower bound for
$1/R$ and $m_D=1$~keV ($100$~keV) gives
$m_{\nu_R}\gtrsim 170$~GeV (40~GeV) so that an order $10$~GeV $\nu_R$
is disallowed. For masses $m_{\nu_R}\ge
10^2$~GeV the bound (\ref{heavy_nu_bound_bulk_ql_2}) may be
satisfied and thus consideration of the radiative decay of $\nu_R$
into light neutrinos in a cosmological context leads to the bound
$m_{\nu_R}\ge 10^2$~GeV. As mentioned already, the neutrino $\nu_R$
may also decay via
emission of an $SU_l(2)\subset SU_l(3)$ glueball in QL symmetric
models. Before concluding that cosmological considerations
demand $m_{\nu_R}\ge 10^2$~GeV we must further investigate this decay
mode.
\subsection{Glueball Mediated Heavy Neutrino Decays\label{bulk_ql_glueball_nu_decay}}
The interaction eigenstate right-chiral neutrino
is an $SU_l(2)$ singlet and as such doesn't couple
directly to the
$SU_l(2)$ gluons (henceforth referred to as `stickons', adopting the
notation of~\cite{Babu:2003nw}). However the
physical mass eigenstate $\nu_R$ does couple to the stickons. Recall
that $SU_l(2)$ is predicted to be an unbroken symmetry and is thus
expected to be confining. Stickon exchange leads to bound state
fermions
formed by the liptons and the $Y$ bosons. Interestingly these bound state
fermions possess the same quantum numbers as the SM
leptons~\cite{Foot:1992sv}. Consequently they mix with the known
leptons and the resulting physical leptons contain small admixtures of states
which couple to the stickons.

Of importance to us is the mixing of $\nu_R$ with the SM singlet bound
state fermions (henceforth bound state neutrinos). It has been shown
that the neutrino-bound state neutrino mixing leads to heavy neutrino
decays of the type $\nu_R\rightarrow G \nu_L$, where $G$ is an
$SU_l(2)$ glueball (henceforth a stickball). The stickballs have a
mass set by the
$SU_l(2)$ strong interaction scale $\Lambda_{SU_l(2)}$, which has been
  estimated in~\cite{Foot:1992sv} to be of order 10~MeV. The lightest
  stickball is expected to be a Lorentz scalar $G_s$ and the neutrino
  $\nu_R$ may decay by emitting a real stickball if
  $m_{\nu_R}>\Lambda_{SU_l(2)}$. If $m_{\nu_R}<\Lambda_{SU_l(2)}$ the
  stickball must be virtual and the decay mode is $\nu_R\rightarrow 3\nu_L$.

The mixing between $\nu_R$ and the bound state neutrinos was
quantified in~\cite{Foot:1992sv} in terms of two mixing angles
$\theta_{1,2}$. It was found that a maximal value of $\sin^2
\theta_{1,2}\sim 10^{-6}$ is expected for 4D QL models. The
rate for $\nu_R\rightarrow G_s \nu_L$ then depends on
$\sin^4\theta_{1,2}$.

In the 5D theory additional factors arise. Again, because $Y$ boson
exchange only couples $\nu_R$ to $n>0$ KK mode liptons a factor of
$K_1^2$ must be included. Also the study of~\cite{Foot:1992sv} assumed
Dirac neutrinos. We have Majorana neutrinos and the decay rate for
stickball emission will depend on the mixing between $\nu_L$ and
$\nu_R^c$, which
takes the standard seesaw form of $\sin \theta_s\sim m_D/m_{\nu_R}$ in
our model. Putting this together we find that the replacement
\begin{eqnarray}
\sin^2\theta_{1,2}&\rightarrow&K_1^2\sin^2\theta_{1,2}
\left(\frac{m_D}{m_{\nu_R}}\right)\label{bulk_ql_glueball_decay_replacement},
\end{eqnarray}
is required to utilise the results of the 4D
study. In~\cite{Foot:1992sv} it was found that the lifetime for a
17~keV neutrino to decay by stickball emission could be
$\le 10^{5}s$. Employing the replacement
(\ref{bulk_ql_glueball_decay_replacement}) one finds that the lifetime
for stickball emission is $\le 10^{25}s$ for a heavy neutrino with mass
$m_{\nu_R}=10^2$~MeV in the 5D model. The lifetime is inversely proportional to
$m_{\nu_R}$ so that larger mass values decrease the lifetime. However
the dominant decay mode will be $\nu_R\rightarrow \nu_L \gamma$ for
$m_{\nu_R}>10^2$~MeV, so that the bound of $m_{\nu_R}\ge 10^2$~GeV
still applies.

Interestingly both the stickball and photon decay modes for $\nu_R$
depend on the Dirac mass coupling $\nu_L$ and $\nu_R$. For
$m_D\rightarrow0$ these decays do not occur and $\nu_R$
becomes stable. In this limit it is important to ensure
that the density of $\nu_R$ particles which remain at the freeze out
temperature of $T_*\sim m_{\nu_R}/15$ does
not overclose the universe. This leads to the well known Lee-Weinberg
bound of $m_{\nu_R}\ge 5$~GeV for a massive Majorana neutrino. This is
much weaker than the bound obtained when considering the radiative
decay. In the present model the limit $m_D\rightarrow0$ corresponds to
the limit in which $\nu_R$ is localised at the opposite boundary to
$\nu_L$. We note that this limit does not lead to rapid proton decays
of the type $p\rightarrow\nu_R\pi$ as kinematic considerations of any neutrino which satisfies the
Lee-Weinberg bound will preclude proton decay via this channel.  Given
that this is the lowest bound on the mass of $\nu_R$ we will consider
this scenario in what remains. This setup has the added advantage of
allowing for an improved understanding of the hierarchy between quark
and lepton masses in a QL symmetric framework, a matter which is
currently under
investigation~\cite{5D_ql_flavour_study}.
\section{Bounds On The Remaining Scales And Experimental Signatures\label{sec:bulk_ql_bounding_scales}}
Having obtained the lower bound on the mass of $\nu_R$ in the
preceding section we may now consider the bounds on the remaining
scales in the theory. The QL symmetry breaking scale must satisfy
the bound $w\gtrsim 1$~TeV, due to leptonic annihilations involving the
$Z'$ bosons. If we take $1/R=30$~TeV in accordance with the lower
bound obtained by considering KK contributions to FCNCs, then the lower
bound on $m_{\nu_R}$ and the upper bound of 1~eV on $m_{\nu_L}$
translate into lower bounds on $w$ and $\Lambda$. Using
(\ref{bulk_ql_sm_neutrin_mass}) and (\ref{bulk_ql_sterile_nu_mass})
gives:
\begin{eqnarray}
w&=&\left(\frac{u^4 m_{\nu_R}^5\pi
    R}{m_{\nu_L}^2}\right)^{1/6},\nonumber\\
\Lambda &=& \left(\frac{m_{\nu_R}u^2}{m_{\nu_L}\pi R}\right)^{1/3},\nonumber
\end{eqnarray}
so that the lower bound on $m_{\nu_R}$ translates into the bounds
\begin{eqnarray}
w&\gtrsim& 5~\mathrm{TeV},\nonumber\\
\Lambda&\gtrsim&460~\mathrm{TeV},\nonumber
\end{eqnarray}
for $1/R=30$~TeV. Furthermore the demand that the proton does not
decay too rapidly translates into a bound on $v$, which we find to be
$v\gtrsim 420$~TeV. This enables us to summarise the mass scales for
the exotic fields in our construction and some of the associated phenomenology.

With $1/R=30$~TeV and $m_{\nu_R}=O(\mathrm{GeV})$ in
accordance with the lower bounds, we find the zero mode $Z'$ boson
mass to be $M_{Z'}^{(0)}=5$~TeV. The heaviest liptons acquire an
order $w$ mass and all liptons will appear at energies $\lesssim
w$. These liptons do
not couple directly to the known leptons via $Y$ exchange. They
appear in different $SU_l(3)$ multiplets and the $Z_2\times Z_2'$
parities preclude a direct coupling of the type $YL_1L_2$. This is a
major distinction
between this model and previous QL symmetric constructs.

Assuming order one Yukawa couplings the lightest liptons will possess masses less than $w$. The lower
bound of $w\gtrsim 5$~TeV permits these liptons
to appear at TeV energies and thus these states may be observed at the
LHC. Although they do not couple directly to leptons via $Y$ boson
exchange they will couple with the known fermions through
electroweak interactions and through $Z'$ exchange. Thus the $n=0$
liptons may appear at the LHC via interactions of the type
$p\bar{p}\rightarrow \gamma, Z \rightarrow L\bar{L}$ etc. A key
signature for this construct would be the appearance of liptons at the
LHC. One could then discriminate this model from other QL symmetric models
by studying the coupling of the liptons to leptons at the proposed
ILC. The liptons of this model do not couple to leptons via $Y$ boson
exchange and thus only electroweak and $Z'$ interactions would couple
$e^+e^-$ pairs to the liptons. 

We remind the
reader that the unbroken $SU_l(2)$ symmetry confines liptons into two
particle bound states. Having the lightest liptons in different
$SU_l(3)$ multiplets to the SM leptons does not alter the stability of
the lightest lipton bound states. These decay via the
electromagnetic or weak interactions. As the electroweak bosons have
uniform profiles in the fifth dimension their couplings to the liptons are
the same as the 4D case. Thus the lifetimes of the lightest confined
liptons are the same as the 4D case and these bound states present no
cosmological concern.

The liptons
which do couple to the SM leptons are $n>0$ members of KK
towers. These states possess mass of order $v\sim 420$~TeV. Their
couplings to SM leptons via $Y$ exchange are highly suppressed due to
the localisation methods employed in this work, thus the associated
phenomenology (like the rare decay $\mu\rightarrow 3e$) is also suppressed. 

Whilst new physics may appear at TeV energies in this model, it
is not until energies of order $30$~TeV that the higher dimensional
nature of the theory reveals itself. At these scales the KK
excitations for the neutral gauge bosons $Z$ and $Z'$, the photon, the gluons
the $W$ bosons appear. The $Y$ bosons also appear at this scale,
though at these energies they will only manifest themselves in
precision experiments through
couplings to the other gauge bosons.
\section{Concluding Remarks\label{sec:bulk_5d_ql_conc}}
We have constructed a complete five dimensional QL symmetric
model. This model differs from a previous five dimensional QL
symmetric model~\cite{McDonald:2006dy} in that all fermions are assumed to propagate in the
bulk. Placing fermions in the bulk provides the following advantages
advantages over the earlier framework. Namely:\\
$\bullet$ The longevity of the
proton is readily understood by localising
quarks and leptons at (or near) different fixed points
in the extra dimension.\\
$\bullet$ The extra dimension may be as large as $R$=1/(30~TeV) allowing the phenomenology associated with
the KK towers of the gauge sector to be observed at
future colliders.\\
$\bullet$ The higher dimensional framework allows one
to understand the absence of mass relations of the type
$m_e=m_u$ or
$m_e=m_d$ in a QL symmetric framework due to the different profiles of
quark and lepton wavefunctions in the extra dimension.\\
$\bullet$ The five dimensional
model permits a purely higher dimensional mechanism whereby one may
suppress
the neutrino mass scale relative to the electroweak scale by spatially
separating left- and right-chiral neutrino fields.

The bounds on the mass of $\nu_R$ depend on its localisation point. We
find that $m_{\nu_R}\ge100$~GeV is required if $\nu_R$ is localised
near $\nu_L$, whilst $\nu_R$ can be of order GeV if it is localised at
the `quark end' of the extra dimension.

The model as it stands has some features which are introduced in a
somewhat arbitrary fashion and it would be pleasing to uncover deeper
reasons for their implementation. In particular it would be satisfying
to understand why
the five dimensional fermion $L_1$ couples more strongly to the bulk
scalar $\Sigma$ than do $N_{R1}^c$ and $E_{R1}^c$. It would
also be pleasing to discover a connection between the fermions that
undergo $SU(3)$ symmetry breaking and the fermions which couple with
opposite sign Yukawa coupling constants to the bulk scalars
$\Sigma$ and $\sigma$. In four dimensional QL symmetric models one
starts with two sets of fermions which are indistinguishable at high
energies. At low energies leptons are defined as those fermions which
experience $SU(3)$ symmetry breaking. In our five dimensional model
there are two independent features which distinguish quarks from
leptons. Leptons are defined to be those fermions which experience
$SU(3)$ symmetry breaking \emph{and} couple to $\sigma$ and $\Sigma$
with opposite signs. It would be pleasing to develop a mechanism which
ensures that the fermions which undergo $SU(3)$ symmetry breaking \emph{must} also
couple to the two bulk scalars with opposite signs.

We note also that some interesting steps
towards understanding flavour in a five dimensional Left-Right
symmetric model have been made in~\cite{Hung:2002qp}. An intriguing
direction for further study would be to attempt to combine the methods
employed in~\cite{Hung:2002qp}
with those of the present study.
\section*{Acknowledgements}
The authors thank Ray Volkas, Alison Demaria and Damien George for
discussions and A. Perez-Lorenzana for a helpful correspondence. This
work was supported in part by the Australian Research Council. 

\end{document}